**Optical spin orientation of minority holes in a modulation-doped GaAs/(Ga,Al)As quantum well**

A.V. Koudinov[1,2], R.I. Dzhioev[1], V.L. Korenev[1], V.F. Sapega[1], Yu.G. Kusrayev[1]

1 A.F. Ioffe Physico-Technical Institute of RAS, 194021 St.-Petersburg, Russia
2 Spin Optics Laboratory, St.-Petersburg State University, 198504 St.-Petersburg, Russia

The optical spin orientation effect in a GaAs/(Ga,Al)As quantum well containing a high-mobility 2D electron gas was found to be due to spin-polarized minority carriers, the holes. The observed oscillations of both the intensity and polarization of the photoluminescence in a magnetic field are well described in a model whose main elements are resonant absorption of the exciting light by the Landau levels and mixing of the heavy- and light-hole subbands. After subtraction of these effects, the observed influence of magnetic fields on the spin polarization can be well interpreted by a standard approach of the optical orientation method. The spin relaxation of holes is controlled by the Dyakonov-Perel' mechanism. Deceleration of the spin relaxation by the magnetic field occurs through the Ivchenko mechanism – due to the cyclotron motion of holes. Mobility of holes was found to be two orders of magnitude smaller than that of electrons, being determined by the scattering of holes upon the electron gas.

**1. Introduction**

Absorption of the circularly polarized light by a semiconductor leads, due to the angular momentum conservation, to the generation of spin-oriented electrons and holes.[1] Subsequently, the electrons and the holes may lose their nonequilibrium spin by transferring it to third bodies, e.g., to lattice nuclei, magnetic impurities, etc. If the nonequilibrium spin polarization in the electron-hole system is retained long enough, it can be experimentally detected and studied by, most commonly, the polarized luminescence or the polarized transmission/reflection (variants of the pump-probe method).[2] Kinetics of the spin phenomena is closely related with the orbital motion of the charge carriers, kinetics of their excitation and de-excitation, with the crystal symmetry, the band structure, with localized and impurity states. Investigation of these phenomena in aggregate *by the action on the spin* is the essence of the optical orientation method.[1]

In a bulk direct-gap cubic semiconductor like GaAs, the top of the valence band in the $\Gamma$ point is fourfold degenerate, which leads to a rapid loss of the nonequlibrium spin of holes on a timescale of the momentum relaxation time, $\tau_p$. That is why a variety of optical orientation phenomena in such materials (and in many of the derivative nanostructures) is related, mostly, to the nonequilibrium spin of the conduction-band electrons. Optical orientation of holes is nearly absent in bulk cubic semiconductors[3,4,5] and in quantum wells at high energies of the exciting photons,[6,7,8] where the heavy-light hole mixing substantially accelerate the hole spin flips.

At the same time, in nanostructures, in strained cubic and hexagonal crystals where the degeneracy of the heavy and light hole subbands is removed, the spin relaxation of holes (possessing a small enough energy) dramatically slows down.[9,10] In the corresponding bulk crystals, the spin orientation of holes was observed even in the continuous-wave experiments.[1,11] In quantum wells, the hole subband degeneracy is also removed, leading to the optical orientation of holes both at the stationary excitation conditions,[7,12,13] and in the time-resolved experiments.[6,14] In quantum dots, within excitons the optical orientation of holes is usually suppressed by the exchange coupling of the hole with the electron,[15] but within negative trions it was observed in different objects.[16],[17],[18]

In the present paper, the effects of the optical spin orientation are reported for a modulation-doped (001)-GaAs/(Ga,Al)As quantum well (QW) containing a high-mobility 2D electron gas. We show that the observed signal of the photoluminescence (PL) polarization should be ascribed to the nonequilibrium spin polarization of minority charge carriers, the holes. Oscillations of the optical orientation signal in the external magnetic field – up to the "optical orientation sign change" in strong fields – can be well described within a model whose main elements are the resonant absorption of the laser radiation by the Landau levels and the mixing of the heavy and light hole subbands by the in-plane anisotropy of the QW. After subtraction of these effects, the observed influence of magnetic fields on the PL polarization can be well interpreted by a standard approach of the optical orientation method. The spin relaxation of the minority holes is controlled by the Dyakonov-Perel' mechanism.

**2. Sample and experimental details**

The sample studied here has been under investigation in Refs.19, 20. It contained a single modulation-doped GaAs QW with a 2DEG concentration $n_e \approx 1.8 \cdot 10^{11}$ cm$^{-2}$. The structure was grown on a (001)-GaAs substrate followed by a thick GaAs buffer layer, the 100-period GaAs/Ga$_{0.7}$Al$_{0.3}$As superlattice designed for the strain relaxation, then by a 250 Å wide quantum well of GaAs, then by a thick Ga$_{0.7}$Al$_{0.3}$As layer which included the $\delta$-doped (Si) layer; the sequence was finalized by the GaAs cap layer. The 2DEG in the quantum well was characterized by a high value of the Hall mobility, $\mu_e > 2 \cdot 10^6$ cm$^2$/Vs at 2 K, which evidenced the high technological quality.[21]

Optical experiments were held with the sample immersed in the liquid helium at 2 K. Measurements of spectra and of the polarization degree of the PL in weak magnetic fields, as well as of the Hanle effect in strong fields were performed using double spectrometers and photomultipliers; the polarization-resolved measurements with a high spectral resolution were held using a DILOR XY800 spectrograph equipped by the CCD detector.

Measurements of the polarization degree were accomplished using the quartz photoelastic modulator (PEM) and the two-channel photon counting. In the optical orientation experiments, we measured the circular polarization degree $P_{Circ} = (I_+ - I_-)/(I_+ + I_-)$, where, by definition, $I_+$ stands for the intensity of the circular component of the PL polarized as the laser light, $I_-$ is the intensity of the opposite circular component. Usually, the laser excitation had a fixed circular polarization, while collection of the PL into the two polarization channels was performed through the PEM. Sometimes we applied a configuration with the alternating polarization of the laser but the fixed recorded polarization (e.g., in order to check if the nuclear spin polarization effects could influence our results). Either measurement configuration had its well-known peculiarities:

for instance, in the first scheme, the signal has a contribution from the equilibrium circular polarization of the PL, while in the second scheme, from the absorption circular dichroism. Finally, we also measured the equilibrium (more correct, non excitation-polarization dependent) linear polarization of the PL. The linear polarization degree was defined as $P_{Lin} = (I_{110} - I_{1\bar{1}0})/(I_{110} + I_{1\bar{1}0})$, where $I_{110}$ and $I_{1\bar{1}0}$ denote the PL intensities at the orientations of the analyzer along the corresponding crystallographic directions.

## 3. Experimental results

### 3A. Zero external field

A typical PL spectrum of the QW excited by the laser with the energy of quanta well above the PL energies ("the high-energy excitation") is depicted in Fig.1 (a). The full width of the line at a half maximum level comprised ~3.5 meV. According to previous studies, for high-quality wide GaAs QWs the concentration $n_e \approx 2 \cdot 10^{11}$ cm$^{-2}$ should be described as a Fermi sea. [22,23,24] This is not the case, e.g., for CdTe-based QWs where trion-like multi-particle states may still reveal themselves due to a smaller exciton radius in those compounds.[25, 26]

Fig 1 (a) shows also the PL excitation (PLE) spectrum taken with the spectrometer position at the PL maximum. The lowest-energy PLE maximum is ~4 meV away from the maximum of the PL; also the PLE maximum at ~7 meV above the PL peak is observed. We attribute the first PLE maximum to the radiative transitions (EHH) from the lowest-energy hole subband, the heavy holes, to the conduction band. The second PLE maximum corresponds to the transitions from the light-hole subband (ELH). The significant Stokes shift between the PLE and PL maxima reflects the Moss-Burstein shift.

Before we pass on to the quasi-resonant excitation of the PL, where our attention will be mainly on the polarization measurements, we should note that at high-energy excitation, a quite significant *linear* polarization of the PL was observed (the polarization degree $P_{Lin} \sim 0.1$). The direction of the predominant polarization of the oscillations of the electric vector was parallel to the [110] crystal axis, while the polarization degree changed only weakly across the PL spectrum and was independent of the polarization of the laser and of the energy of the laser photons. This "built-in" linear polarization of the PL was an intrinsic property of the sample,[27] being a characteristic of the anisotropy of the photon-emitting states in the plane of the QW.

It has been established by multiple studies that during the epitaxial growth of nanostructures based on cubic semiconductors, the tetragonal in-plane symmetry frequently becomes broken, typically leading to the non-equivalence of the [110] and [1$\bar{1}$0] directions. Microscopic reasons for the broken symmetry can be, for instance, a uniaxial deformation in the plane of the QW layer or a particular role played by one of the QW/barrier heterointerfaces (each the interface has a lower symmetry than the QW in aggregate, including two ideal heterointerfaces). In our case, the latter mechanism can possibly be very efficient in view of the results of Ref.19. It was shown therein that, because of the one-side $\delta$-doping, a built-in electric field was formed in our QW. This electric field is directed along the growth axis, clasping electrons and holes to the opposite interfaces. The valence-band mixing on the low-symmetry interface leads to a partial linear polarization of the PL, with the predominant direction specified by the directions of chemical bonds at the interface.

If the QW was excited by the circularly polarized light with the energy about the EHH (quasi-resonant excitation), the PL turned out to be partially circularly polarized. This polarization signal was due to the optical spin orientation, as it reversed sign in response to the sign reversal of the incident polarization. One can see from Fig.1 (b) that the optical orientation signal is nearly constant across the PL contour. However, it sharply changes (decreases) when the energy of the laser photons is increased (see the excitation spectrum of the polarization in Fig.1 (a)). As soon as the laser excitation extends to the ELH transitions, the PL polarization vanishes.

*3B. Longitudinal magnetic field*

Let us consider first how the optical response of the QW is influenced by a weak magnetic field applied along the growth axis. Fig.2 (a) shows variations of the polarization signal against the increasing field value at fixed energies of excitation and detection. A monotonous increase of the polarization degree at both signs of the applied field is the main trend. Small sub-Tesla values of the field in Fig.2 (a) and parity of the effect in the magnetic field allow one to exclude any noticeable role of the contributions to the circular polarization which are induced by the magnetic field and do not depend on the laser polarization. The same is confirmed by the absence of the PL circular polarization at the linearly polarized laser (not shown). One may conclude that the increase of the polarization degree in Fig. 2 (a) is the increase of the optical orientation signal. This is an expected result, since it corresponds to one of the main paradigms of the optical orientation method – deceleration of the spin relaxation by a longitudinal field.[1]

We also note small oscillation-type anomalies which appear on the field dependences in fields $B \geq 0.2$ T. Looking for the origin of these oscillations, we discovered that a much more pronounced oscillating behavior is demonstrated in the same conditions by the signal of full PL intensity (Fig.2 (a)). We shall show in Section 4A that both the oscillation effects have the same reason: a modulation of the absorption coefficient at the energy of laser excitation by the Landau levels.[28] Therefore it is instructive to study the behavior of the optical response in the high-fields domain, where the Landau quantization should manifest itself stronger.

Fig.3 presents pairs of the PL spectra, corresponding to the two measured circular polarizations, at several values of the applied magnetic field. One can see that in strong fields, the PL spectrum fall to separate narrow lines corresponding to the transitions between the electron and hole Landau levels with appropriate numbers. As the field is increased, the Landau levels move upward in energy, and the energies of the electron-hole pair optical transitions exceed, one by one, the fixed energy of the exciting photons. One can follow in Fig.3 how the Landau-level components of the PL, taking turns, cross the spectral position of the laser. The movement of the Landau levels in the magnetic field can be characterized by the "fan diagram" in Fig.4, showing the positions of the maxima of the PL components versus the field value.

Starting from the fields $B \geq 1$ T, one can trace in Fig.3 shifts between the counter-polarized PL lines. This shift corresponds to the spin splitting of the Landau levels. That is why for strong fields, the pairs of PL spectra taken in two opposite circular polarizations can be more conveniently analyzed than the derivative value of the polarization degree (which would include a contribution due to a spectral shift between the lines). By an inspection of the spectra in Fig.3, it is easy to understand that this series of spectra was taken with the polarization of the laser corresponding to the upper spin sublevels of the Landau levels. To this end, one should recognize that in strong fields, the upper spin sublevels are observed in the same polarization that provided the larger PL intensity in the zero field due to the optical orientation. We have also measured the analogous series of spectra for the opposite polarization of the laser, where the lower spin sublevels of the Landau levels were excited (not shown).

In Fig.3, starting from the zero field, the optical orientation effect manifests itself in a larger intensity of the entire spectrum of the co-polarized (with the laser) PL component than that of the counter-polarized PL component (Fig.3 (a-d)). However, as the field is increased, the normal intensity ratio suddenly changes to the opposite one (Fig. (e)), while at a further increase of the field the normal situation is restored (Fig.3 (f)). We note that a similar anomaly was observed with the opposite laser polarization (this series of spectra is not shown). Here the anomaly was shifted to higher fields and manifested itself in the conditions similar to those of Fig.3 (f). We shall see in Section 4B that the nature of both anomalies is the same as that of the oscillation behavior of the PL intensity and polarization in weak magnetic fields.

*3C. Transverse magnetic field*

The field applied in the QW plain (the transverse field) does not cause so dramatic modifications of the PL spectra, as the QW confinement hinders the cyclotron motion of the charge carriers, and no Landau levels are formed. Still the transverse field makes an effect on the PL polarization. One can see in Fig.5 that a suppression of the optical orientation by the field – the Hanle effect – is observed. The Hanle dependence is well fitted by a Lorentzian contour with half-width 3.7 T. A 1 meV shift of the PL maximum by the in-plane magnetic field,[29] is not important for polarization measurement, because the polarization degree does not vary over PL spectrum (see Fig.1 (b)).

One should note an unusually large value of the characteristic (transverse) field of the Hanle effect as compared to the characteristic (longitudinal) field of the slowdown of the spin relaxation. A typical situation for the optical orientation in semiconductors is that the Hanle effect occurs in weak fields but the spin relaxation slowdown – in stronger fields. There exist special cases where these two characteristic fields are close in value, being controlled by a same factor.[30] However, in our measurements the Hanle contour half-width (3.7 T) is an order of magnitude larger than the characteristic field of the deceleration of the optical orientation (~0.25 T, see Fig.2 (a)).

**4. Discussion of the results**

*4A. Identification of the optical orientation of holes*

In many semiconductor systems, the optical orientation effect is due to the spin memory of conduction-band electrons, while the spin memory of holes often does not last sufficiently long. We have been led to the idea about the hole-related origin of the observed optical orientation signal by the large half-width of the Hanle curve. Indeed, within a standard approach the Hanle effect contour is[1]

$$P_{Circ}(B) = \frac{T_s}{\tau} \cdot \frac{1}{1 + \left(\mu_B g T_s \hbar^{-1} B\right)^2}, \qquad (1)$$

where $g$ is the effective g-factor of a spin-polarized particle, $T_s$ is the lifetime of the nonequilibrium spin ($T_s^{-1} = \tau_s^{-1} + \tau^{-1}$, $\tau_s$ and $\tau$ – spin relaxation time and the particle lifetime, respectively), $\mu_B$ – the Bohr magneton. The width of the contour Eq.(1) is the more, the less is the product $gT_s$. However the time $T_s$ cannot be too small as compared to $\tau$, since the ratio $T_s/\tau$ constrains the amplitude of the contour (the zero-filed value). Hence the observed large contour half-width makes one to allow for either an unreasonably small lifetime of a particle in a QW (at an even shorter spin relaxation time) or, more reasonably, a small value of the g-factor of the particle. The latter can be easily grounded if the particle carrying the nonequlibrium spin is the hole. The point is that the intrinsic transverse g-factor of the heavy hole in a QW is very small – of an order $10^{-2}$. Frequently, in reality the transverse g-factor value of ground-state holes is determined by contributions related to the breakdown of the in-plane symmetry. If this is the case (like, we shall see, in our QW), the characteristic value of the transverse g-factor of holes can be ~0.1, i.e., still noticeably less than that of electrons (0.44).[31]

The second argument for the spin polarization of holes is the spectrum of the optical orientation. In case of a degenerate electron gas, the spin polarization of electrons is possible only in a narrow energy interval corresponding to the thermal smearing of the Fermi distribution, while at other energies the numbers of spin-up and spin-down electrons are equal (the 2D density of states does not depend on energy). Since the entire 2DEG participates in the formation of the PL spectrum, it should have resulted in a sharp spectral dependence of the optical orientation.[32] However in the experiment, the optical orientation practically does not change across the PL spectrum (Fig.1 (b)).

The third argument is a sharp dependence of the optical orientation across the PLE spectrum (Fig.1 (a)). As discussed in Section 3A, the polarization signal immediately vanish when the energy of the laser quanta exceeds the EHH transition, and the ELH transition become excited. Disappearance of the polarization can be most naturally ascribed to the change of the involved hole state. Previous studies have also shown that in case of the optical orientation of holes, the dramatic decrease of the signal typically takes place when the exciting laser is tuned off the quasi-resonant energy positions.[7,12,33]

*4B. Oscillation behavior and anomalies in the polarization*

This section is devoted to the elucidation of the nature of the oscillation behavior of the PL intensity and polarization in weak magnetic fields (Fig.2 (a)) and of the anomalous sign reversals of the optical orientation in strong fields (e.g., Fig.3 (e)). We shall see that both effects have one source. Since the main task of the present paper is a study of the optical orientation effect, we shall not look for a too detailed quantitative description of the anomalies but will restrain ourselves by a discussion of their reasons and by an illustrative calculation.

Weak oscillations of the PL polarization in Fig.2 (a) are seen together with the more pronounced oscillations of the full PL intensity, which will be a starting point of our analysis. Assume, the oscillations of intensity are caused by resonance absorption of light at the Landau levels.[28,34] As the magnetic field is increased, the Landau levels cross, one by one, the spectral position of the laser, causing flashes of the PL intensity due to the bursts of the number of the created electron-hole pairs. The flashes are not seen in weak fields where the Landau levels are located tight in energy, each of them having a small capacity. But at stronger fields, the separation of the levels increases, the levels with smaller numbers come to the resonance with the laser, and the oscillations show up.

Parameters of this model can be specified using the fan diagram in Fig.4. Here lines represent our fit by an elementary model which describes the positions of all the Landau levels with a single reduced mass and a single effective "bottom of the band" and with no account of the excitonic effects. One can see that, in general, this approximation adequately reproduces the experimental positions of the PL peaks in the intermediate field range, where the peaks become spectrally resolved. Hence the following parameters are determined: the "mobility" of the Landau levels in the magnetic field (a quantum of the cyclotron energy normalized by the field value) ~2.1 meV/T, the position of the "bottom of the band" ~1518.4 meV.

Analysis shows that the oscillations of intensity in Fig.2 (a) are periodic in the inverse field (see inset in Fig.2 (a)) with a period ~0.6 $T^{-1}$. In the framework of the same simplistic model, which even does not account yet for the spin splitting, the period of oscillations in the inverse field should be equal to the aforementioned "mobility" of the Landau levels (2.1 meV/T) divided by the energetic distance from the "bottom of the band" to the laser (3.5 meV). This gives the value 0.6 $T^{-1}$, in perfect agreement with theory. In general, the calculated behavior of the full intensity also well reproduces the experiment (cf. panels (a) and (b) in Fig.2). Here in the calculation, every Landau level was represented by a Lorentzian contour with a single fixed half-width (0.25 meV). Now we additionally accounted for the spin splitting of the Landau levels, which almost does not influence the results so far. We assumed the spin splitting independent of the Landau level number and calibrated its approximate value (~0.1 meV/T) by the spectrally-resolved spin sublevels in strong fields (Fig.3).

So, the oscillations of the PL intensity are reasonably described by the resonant absorption, i.e., by oscillations of the number of the generated electron-hole pairs. But how the oscillation behavior can penetrate in the circular polarization measurement channel, if the polarization degree is intensity-independent? To answer this question, one needs to recall the big value of the "built-in" linear polarization of the PL (Section 3A) and take into account the mixing of heavy and light hole subbands by the anisotropic perturbation in the plane of the QW. The "built-in" polarization, insensitive to the conditions of the laser excitation, will be a first consequence of the mixing. But in the conditions of optical orientation, the valence band mixing also means that every optical transition corresponding to, e.g., excitation of the upper spin sublevel of every Landau level will occur not in a pure sigma plus polarization, but, with a certain probability, in a sigma minus polarization as well (and vice versa – for the lower spin sublevels). Now the incoming light of a fixed circular polarization can generate two types of electron-hole pairs. When exciting "its original" spin sublevel (as without the mixing), the light creates electron-hole pairs which will, most likely, emit photons of the same polarization. When exciting the "alien" spin sublevel for the expense of the valence mixing, the light creates pairs which will finally emit in the opposite polarization with an overwhelming probability. The PL polarization degree will be directly influenced by the ratio in which the two types of pairs are created. This ratio can depend on the laser energy and/or the value of the applied magnetic field.

With no account of the valence mixing, the laser of a specific circular polarization could excite, say, only a "comb" of the upper spin sublevels of the Landau levels. Then the increase of the field would be accompanied by just a modulation of the number of the excited pairs – while their spin polarization would remain at a 100% level. With an account of the mixing, the "comb" of the lower spin sublevels can also be excited (with a smaller efficiency), leading in total to the modulation of the average spin polarization of the excited pairs on the increase of the field. Without any consideration of the details of the spin evolution, but merely by the modulation of the selection rules for the incoming light, one can explain oscillations and even a sign reversal of the optical orientation.

The specific calculation shows that if the model which have been reproducing the behavior of the intensity (Fig.2 (b)) is supplemented by the account of the valence mixing, it satisfactorily explains appearance of the oscillations on the very edge of the weak-field range (inset in Fig.2 (b)). It is worth noting that the calculation, in fact, does not contain much arbitrariness: all its parameters (the half-width of Landau levels (0.25 meV), their spin splitting (0.1 meV/T), their "mobility" in the magnetic field (2.1 meV/T), the energetic gap between the laser and the "bottom of the band" (3.5 meV), the coefficient of admixing of the light holes (0.1)) are more or less well known from the experiment. Possible non-perfect helicity of the circular excitation can contribute to the oscillation effect, in a similar way as the valence band mixing. But the presence of the mixing is guaranteed by the strong "built-in" linear polarization which reveals the in-plane anisotropy of the sample.

However, rudeness of the model used for the description of the Landau levels system is obvious. For example, according to the calculation, in strong fields, in the interval between arrivals of the first and zero-th Landau levels into the resonance with the laser energy, absorption should be practically absent, which is not confirmed by the experiment. This fact seems to be related with a more sophisticated shape of the single-Landau-level absorption line, which may be due to the interaction with phonons or to the Auger processes. In the same spirit, with the Lorentzian elementary contours and the chosen parameters set, our simple model does not allow obtaining an inversion of optical orientation as in Fig.3 (e). The "dips" with a negative excited polarization can be achieved, for instance, by replacement of Lorentzians for Gaussians, but even the use of Gaussians does not quantitatively reproduce positions of the dips in the magnetic field (2 T instead of 2.8 T, see Fig.6). Still the calculation reproduces correctly that for the complementary polarization of the laser, the "dip" occurs at larger field values.

The nature of the negative dips can be understood from simple considerations. The dips are observed (Fig.3 (e)) in the fields interval between arrivals of the first and zero-th Landau levels into the resonance with the position of the laser (about 1.5 T and about 4-5 T, respectively). Absorption in this field range is very weak and is controlled by the "tails" of the Landau levels (responsible for the absorption between resonances). It is the spectral shape of the absorption line tails that determines the inversion of the polarization. (That is why the result of our calculation turned out to be so sensitive to the shape of the elementary contour.) For obtaining a dip with the inversion of optical orientation, it is sufficient that, in certain range of field values, the absorption of light were dominated not by "originally" polarized spin sublevels but rather by an "alien" sublevel (i.e., for the expense of valence mixing).

The outlined reason of the polarization dips in strong fields is quite transparent, thus one can safely wink at minor quantitative problems of the simplistic model. So, we have shown that the modulation of the absorption of the laser radiation by the system of the spin-split Landau levels, together with the mixing of the valence subbands by a low-symmetry in-plane perturbation, gives rise to all the observed polarization anomalies.

*4C. Mechanisms of the spin evolution and estimates of the characteristic times*

Having revealed the reason of the anomalies of the polarization, we are faced to a standard situation with a suppression of optical orientation by the transverse field (the Hanle effect) and its stabilization by the longitudinal field; but the characteristic fields of the two effects are in an unusual relation to each other (see Section 3C). Let us analyze the spin evolution mechanisms and the respective characteristic times on the basis of the experimental data.

As a rule, the dominating mechanism of the spin relaxation of charge carriers in pure semiconductors with no inversion symmetry is the Dyakonov-Perel' (DP) mechanism.[1] It results from the spin-orbit splitting of the band-carrier states which are characterized by a wavevector $\vec{k}$. When a carrier is moving in an arbitrary direction (with an exception of few special ones), its spin is precessing in the effective magnetic field whose value and direction depend on those of $\vec{k}$. Within the short correlation times approximation, the period of precession is much longer than the correlation time of the effective field (the characteristic time between the two consecutive events randomizing $\vec{k}$). The spin relaxation occurs as a result of a chain of such events (collisions) separating the turns of the spin in the effective fields which are both random-value and random-direction.

Suppose, in our QW the spin polarization of holes is limited by the DP mechanism.[35,36] Then, in the longitudinal magnetic field, the deceleration of the spin relaxation should occur via the Ivchenko mechanism: as a result of the motion of holes along the cyclotron orbits, systematically changing the direction of $\vec{k}$ in the intervals between collisions.[1,37] As long as the observed spin polarization is not too weak, $P_{Circ} = T_s / \tau \approx \tau_s / \tau$, it should vary in the longitudinal field by virtue of the dependence $\tau_s(B)$ dictated by the Ivchenko mechanism. To estimate this dependence, we chose the inversed Eq. (51) from Ref.[38]:

$$\tau_s(B) = \tau_s(0) \frac{1}{1 - \Theta_2 (\Omega_c \tau_p)^2}, \qquad (2)$$

where $\Omega_c = eB/m_h c$ is the cyclotron frequency ($m_h$ – the in-plane mass of the hole in the QW), $\tau_p$ stands for the momentum relaxation time of the hole. We chose the value $\Theta_2 \approx 2$ for the $\Theta_2$ factor (its exact value is close to 2 and depends on the mechanism of the elastic scattering), while for the in-plane effective mass we took $m_h \approx 0.1 m_0$, where $m_0$ is the free electron mass. In experiment, in the regime of a quadratic increase, the polarization degree doubles at the field $B_{exp} \approx 0.25$ T. In theory, doubling of $\tau_s(B)$ corresponds to the condition $\Omega_c \tau_p \approx \frac{1}{2}$. This gives the following estimate for $\tau_p$ of holes:

$$\tau_p \approx \frac{1}{2} \cdot \frac{m_h c}{eB_{exp}} \approx 1 \text{ ps}. \qquad (3)$$

The spin relaxation time at zero field, $\tau_s(0)$, can be easily estimated from the observed polarization degree $P_{Circ}(0) \approx \tau_s(0)/\tau \approx 0.1$ and a typical value of the radiative lifetime in a QW $\tau \approx 300$ ps; one obtains $\tau_s(0) \approx 30$ ps. A more reliable way of doing the experiment-based estimation rests upon the Hanle effect. If, as usual, the spin relaxation time is not changed by the transverse field, then, according to Eq. (1), the Hanle contour half-width $B_{1/2} \approx \hbar / \mu_B g_{h\perp} \tau_s(0)$. A transverse g-factor of the hole entering this expression can be evaluated from a typical value of the longitudinal hole g-factor $g_h \approx 1$ using a relationship obtained in Ref.27: $g_{h\perp} / g_h = P_{Lin}$. (This relationship connects two different consequences of the valence-band mixing: the transverse *g*-factor of the ground-state hole and the "built-in" linear polarization of the PL.) Since $P_{Lin} \approx 0.1$ (Section 3A) and $B_{1/2} \approx 3.7$ T (Section 3C), one obtains

$$\tau_s(0) \approx \frac{\hbar}{\mu_B g_h P_{Lin} B_{1/2}} \approx 30 \text{ ps} \qquad (4)$$

in full agreement with the first estimate.

For a theoretical evaluation of the spin relaxation time in the DP mechanism, one can use Eq. (48) from Ref.38 or, more conveniently (and more correctly, in view of the quasi-two-dimentional motion of holes in the QW), the expression from Ref.35 :

$$\frac{1}{\tau_s(0)} = \left(\frac{\beta}{\hbar}\right)^2 k_F \tau^*. \qquad (5)$$

Here $\beta$ is a factor at the $\vec{k}$-linear terms in the Hamiltonian, defined in such a way that the energy difference between the hole states with opposite angular-momentum values $E_{3/2}(k) - E_{-3/2}(k) = 2\beta k$, while $k_F$ and $\tau^*$ are the characteristic wave vector of the hole and the characteristic scattering time. For a non-degenerate quasi-equilibrium gas of holes the value $k_F$ can be taken from the relation $\hbar^2 k_F^2 / 2m_h \approx T$ ($T$ denotes the temperature in energy units). As to the $\tau^*$, the value $\tau_p$ from Eq. (3) can be used for it even with an account of the interaction with the 2DEG, because, in contrast to collisions of particles of the same name, the scattering of holes upon electrons gives a full-fledged contribution to $\tau_p$. Having taken a typical value $\beta \approx 60$ meV·Å from Ref.35, we obtain

$$\tau_s(0) \approx \frac{\hbar^4}{2T\beta^2 m_h \tau_p} \approx 20 \text{ ps}. \qquad (6)$$

It is in a very reasonable agreement with the experimental value 30 ps. We should check that $\Delta\varphi \sim \tau_p \sqrt{2m_h T} \beta / \hbar^2 \approx 0.2 \ll 1$ and $\tau_p \ll \tau_s(0)$, justifying *a posteriori* the use of the short correlation times approximation.

Finally, using the obtained value $\tau_p$ (Eq.(3)) one can estimate the mobility of holes:

$$\mu_h = \frac{e\tau_p}{m_h} \approx 2 \cdot 10^4 \text{ cm}^2/\text{Vs}, \qquad (7)$$

which turns out to be two orders of magnitude smaller than the Hall mobility of electrons. Such a strong difference in mobilities can hardly be related to peculiarities of the scattering of particles upon defects in our high-quality QW. Most likely, this fact means that the low mobility of the minority carriers, the holes, is limited by their scattering upon majority carriers, the electrons.

## 5. Conclusions

We studied the optical spin orientation effects in a GaAs/(Ga,Al)As quantum well containing a high-mobility 2D electron gas. We have shown that the optical orientation signal observed at a quasi-resonant excitation should be ascribed to the non-equilibrium spin polarization of holes. The observed oscillations of the polarization signal in the longitudinal magnetic field – up to the "inversion of optical orientation" in strong fields – were well described in a simple model whose main elements are: (i) resonant absorption of the laser radiation by the Landau levels and (ii) mixing of the heavy and light hole subbands by the in-plane anisotropy of the quantum well. Being refined from these effects, the experimental data regarding the influence of longitudinal and transverse magnetic fields on the PL polarization yield to a coherent interpretation in the framework of the standard approach of the optical orientation method.

We found that the spin relaxation of holes is controlled by the Dyakonov-Perel' mechanism, as previously established for *p*-doped quantum wells.[35] A suppression of the spin relaxation by the longitudinal magnetic field occurs through the Ivchenko mechanism – due to the cyclotron motion of holes. We evaluated the hole spin relaxation time (20–30 ps) and the hole momentum relaxation time (1 ps). Mobility of holes in the quantum well was found to be two orders of magnitude smaller than that of electrons and, seemingly, is determined by the scattering of holes upon the electron gas.

**Acknowledgements**

We are grateful to M.M. Glazov and L.E. Golub for helpful discussions on the issues of the spin relaxation of holes in quantum wells, to B.M. Ashkinadze for providing the previously studied sample [20,24,29] grown by L. N. Pfeiffer[21].

This work was partially supported by RFBR (project 13-02-00316-a), by Samsung 2011 GRO Project and by Russian Ministry of Education and Science (contract No. 11.G34.31.0067 with SPbSU). AK gratefully acknowledges a support from Dmitry Zimin "Dynasty" Foundation.

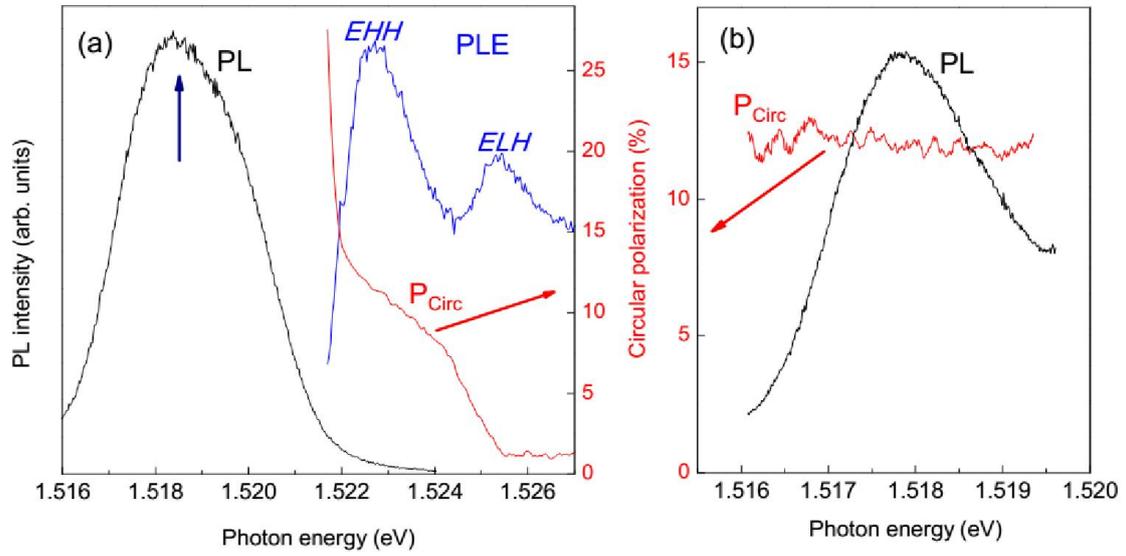

Fig.1 General picture of the optical orientation effect in a modulation-doped *n*-GaAs/(Ga,Al)As quantum well at zero field. (a) The photoluminescence spectrum at a high-energy excitation 1.547 eV (PL); the photoluminescence excitation spectrum (PLE) and the excitation spectrum of the optical orientation ($P_{Circ}$), both taken with a detector spectral position at the energy shown by vertical arrow near the PL contour. (b) The PL spectrum and the optical orientation spectrum taken at the laser excitation 1.5225 eV.

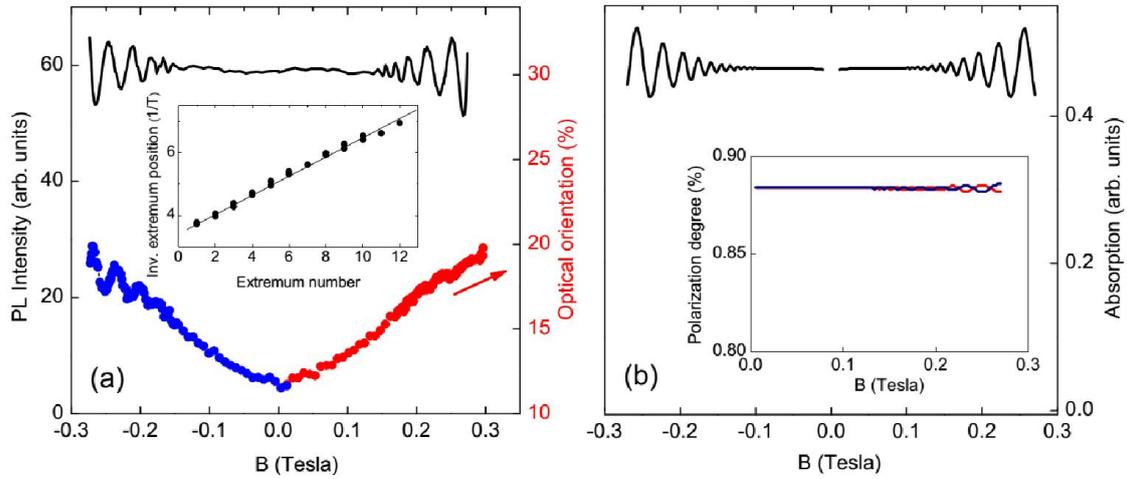

Fig.2 (a) Oscillations of the PL intensity and the behavior of the optical orientation signal in the *n*-GaAs/(Ga,Al)As QW subject to weak longitudinal magnetic fields. Excitation at 1.5220 eV, detection at 1.5185 eV. The inset shows positions of the extrema (maxima and minima) of the oscillating intensity, plotted in the scale of the inverse absolute values of the magnetic field, against the extremum number. The approximating straight line gives a full period of oscillations in the inverse field 0.6 T$^{-1}$. (b) Calculated behavior of the resonant absorption on the Landau levels, explaining the oscillations of the PL intensity. The inset shows the calculated behavior of the mean spin polarization degree of the created holes in the same conditions (the two curves correspond to the two possible circular polarizations of incident light with respect to the magnetic field, cf. with panel (a)). One can see that the polarization anomalies are manifested at stronger fields as compared to the intensity anomalies. See text for the discussion of the parameter values for the calculation.

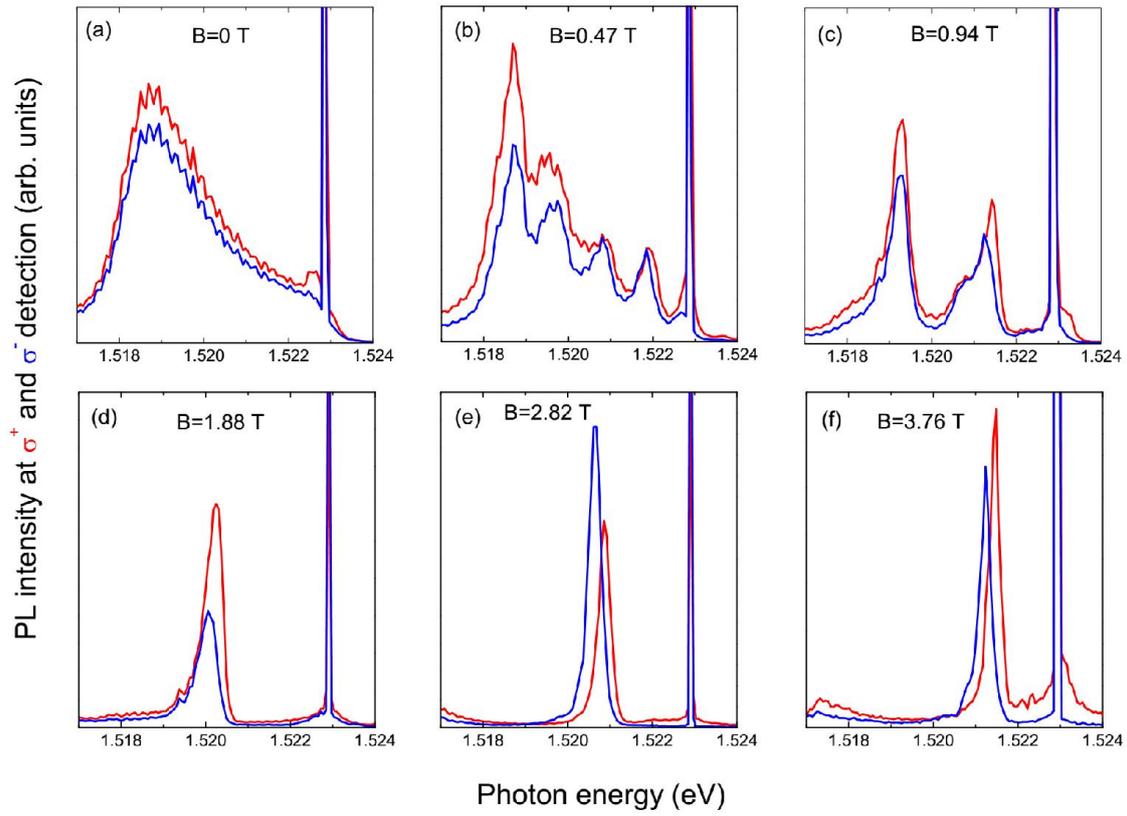

Fig.3 Spectra of two circular components of the PL at moderate and high values of the longitudinal magnetic field and at laser excitation about 1.523 eV (the laser position is seen in the figure). In the magnetic field, the PL falls to the series of optical transitions between the Landau levels, only one of them ($n = 0$) remaining below the laser energy at fields >1 T. The incoming circular polarization (of the laser) corresponds to the predominant excitation of the upper spin sublevels of the Landau levels. The PL spectra taken in the same polarization are shown by red, in the opposite polarization – by blue.

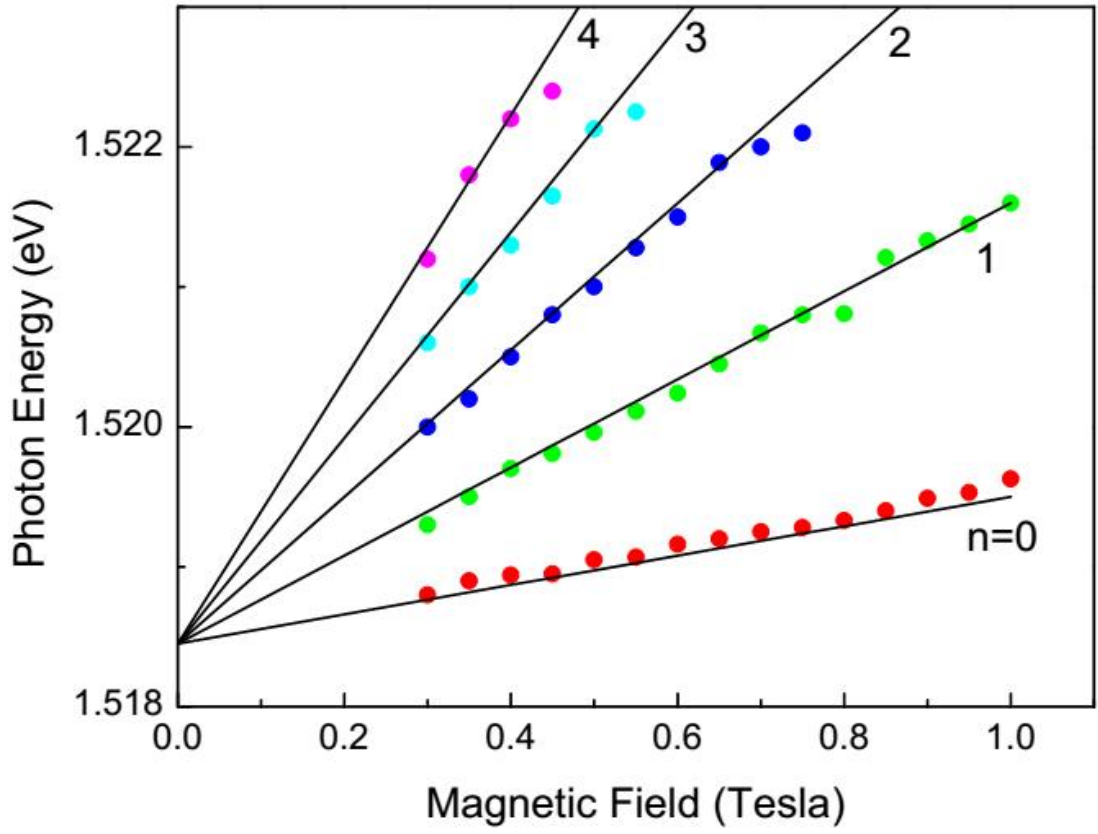

Fig.4 Fan diagram of the positions of the lowest Landau levels (as resolved in the PL spectra at moderate magnetic fields, cf. Fig 3 (b)). Excitation at 1.5229 eV. Straight lines present calculation in the simplest model with the characteristic slope 2.1 meV/T.

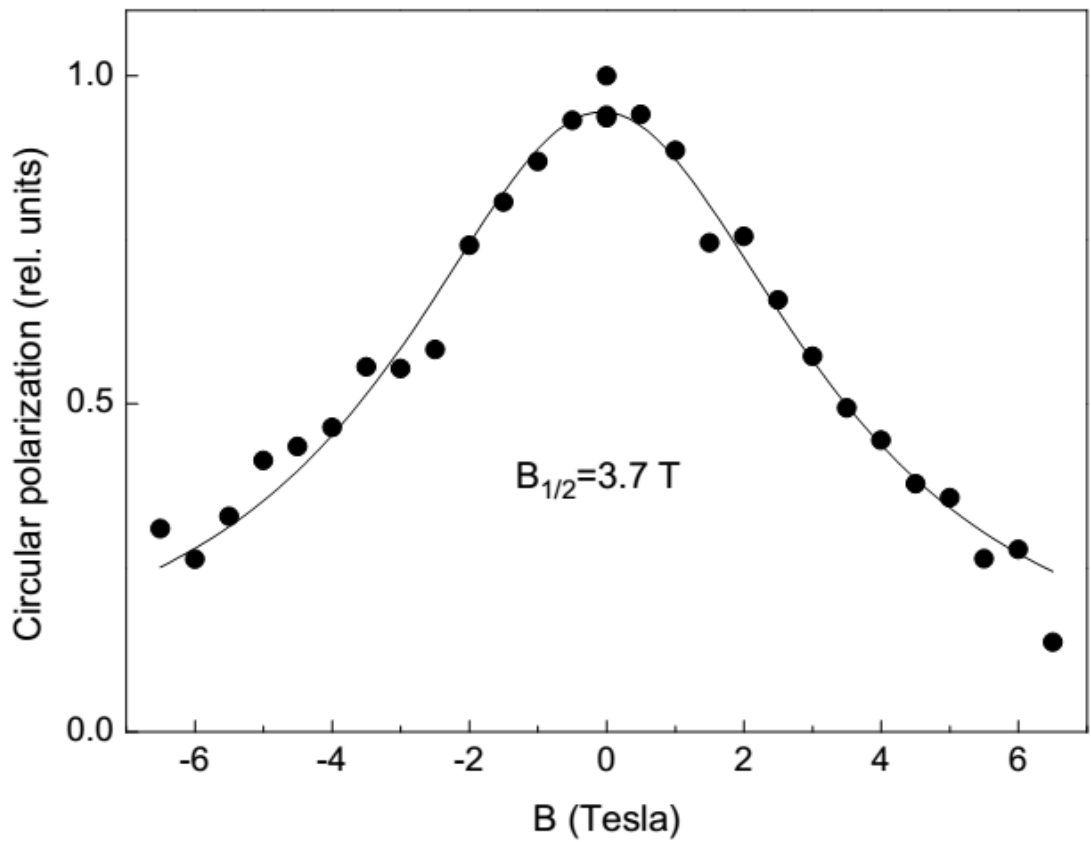

Fig.5 Hanle effect in a *n*-GaAs/(Ga,Al)As quantum well – a suppression of the optical orientation by the transverse magnetic field. The solid curve shows a Lorentzian fit.

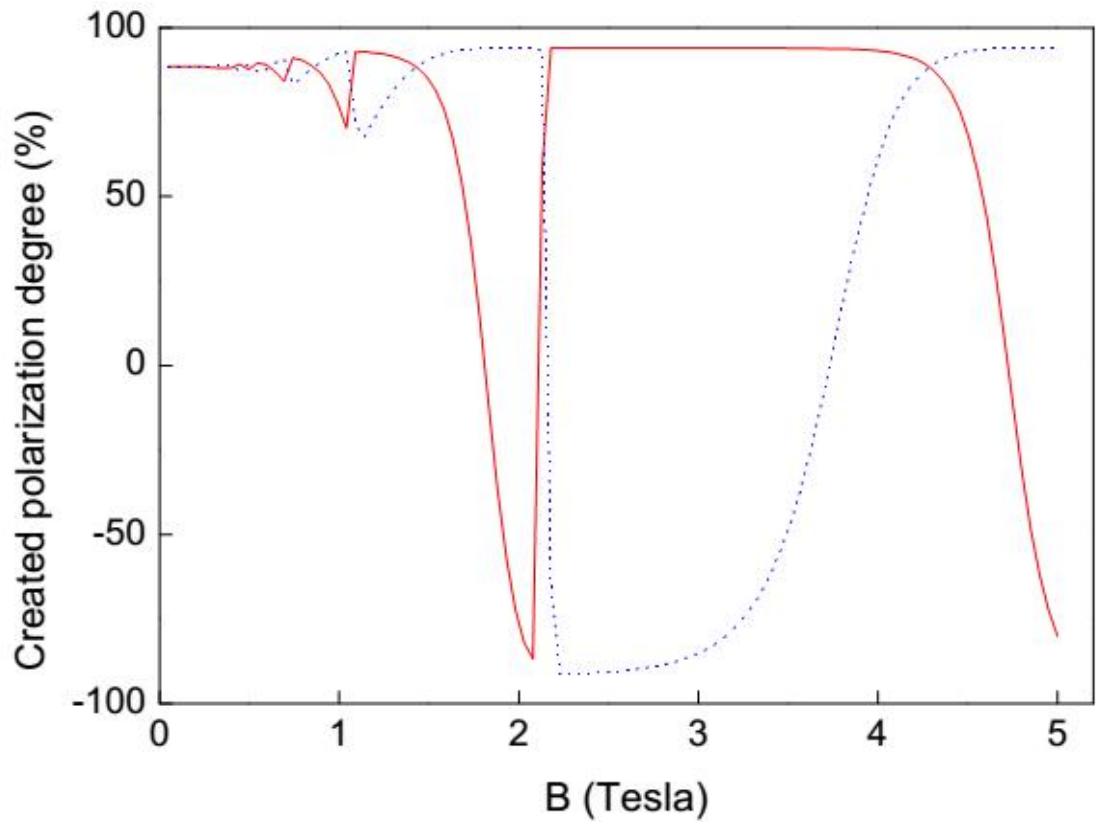

Fig.6 Calculated dependences of the mean polarization degree of the excited spins on the longitudinal magnetic field. Parameters of the calculation as for Fig.2 (b), but the distance between the laser and the "bottom of the band" taken in compliance with conditions of Fig.3. The red curve corresponds to the polarization conditions of Fig.3; a dip around 2 T explains the "inversed optical orientation" in Fig.3 (e). The blue curve corresponds to the opposite polarization of the incident light.